\documentclass[%
 reprint,
 amsmath,amssymb,
 aps,
]{revtex4-2}

\usepackage{graphicx}
\usepackage{dcolumn}
\usepackage{bm}

\usepackage{xcolor}
\begin{document}

\preprint{APS/123-QED}

\title{Gravitational waves in  Ho\v{r}ava-Lifshitz anisotropic gravity}

\author{J. Mestra-Páez}
 \email{jarvin.mestra@ua.cl}
 \affiliation{Departamento de F\'isica, Universidad de Antofagasta, Aptdo 02800, Chile.} 
\author{J.M. Peña}
 \email{joselen@yahoo.com}
 \affiliation{Departamento de F\'isica, Universidad de Antofagasta, Aptdo 02800, Chile.} 
 \author{A. Restuccia}
 \email{alvaro.restuccia@uantof.cl}
 \affiliation{Departamento de F\'isica, Universidad de Antofagasta, Aptdo 02800, Chile.}

\date{\today}

\begin{abstract}
We show that in the anisotropic Ho\v{r}ava-Lifshitz gravity there is a well-defined wave zone where the physical degrees of freedom propagate according to a non-relativistic  linear evolution equation of high order in spatial derivatives, which reduces to the wave equation at low energy. This is so, provided the coupling parameters satisfy some restrictions which we study in detail. They are imposed  to obtain a finite ADM gravitational energy, which depends manifestly on the terms which break the Lorentz symmetry of the formulation. The analysis we perform is beyond the linearized approach and includes all high order terms of the Hamiltonian potential.
\end{abstract}

\maketitle

\section{Introduction}
The Ho\v{r}ava-Lifshitz gravity theory \cite{Horava2009, Blas2010} has been proposed as a candidate of a renormalizable gravity theory \cite{charmousis2009,visser2009,papazoglou2010,orlando2009,shu2009,benedetti2014,contillo2013,d2014asymptotic,d2015covariant, barvinsky2016,BellorinRestuccia2016B,wang2017,shin2017,pospelov2012}. Following Lifshitz, the time and the spatial coordinates scale differently, in a way that the overall coupling of the theory becomes dimensionless. In this sense the theory is anisotropic and non-relativistic. 
It introduces interaction terms with high order spatial derivatives in the potential that
break in a manifest way the relativistic symmetry, but improve the ultraviolet (UV) behaviour in comparison to GR. The theory is renormalizable by power counting.  
There are different versions of the Ho\v{r}ava-Lifshitz gravity, the projectable one with interesting applications to cosmology \cite{Mukohyama2010,jamil2020} and references therein, and the non-projectable one where the full number of gravitational physical degrees of freedom, the transverse-traceless tensorial modes at the linearized  level,  become dynamical.
 Among the non-projectable Ho\v{r}ava-Lifshitz versions the propagating degrees of freedom differ according to the value of the dimensionless couplings $\lambda$ on the kinetic term of the action. For $\lambda \ne 1/3$ the theory propagates, besides the transverse-traceless tensorial degrees of freedom, a scalar one. 
Several works have discussed the problem of strong coupling of this scalar mode \cite{charmousis2009, blass2010, papazoglou2010}.

Some restrictions on the couplings of the theory have to be imposed in order to justify the existence of this scalar mode without violating the well established gravitational data \cite{ramos2019}. 
For $\lambda = 1/3$, the kinetic term on the action has an additional conformal symmetry. In this case the propagating degrees of freedom exactly coincide with transverse-traceless tensorial modes of GR. No additional scalar field is present in the theory. 
At low energies the theory depends on two coupling constants $\beta$ and $\alpha$, when $\beta=1$ and $\alpha=0$ the field equations are exactly the GR equations on a particular gauge.
The restrictions on the coupling constants are in this case less stringent. In both cases there is a range of values for the coupling constants for which the theory fits the known gravitational experimental data satisfied by GR \cite{EmirGumrukcuoglu2018}.

The linearized theory at low energies coincides with the corresponding linearized GR formulation \cite{BellorinRestucciSotomayor2013}. Additionally it satisfies the analogous to the Einstein quadrupole radiation formula \cite{BellorinRestuccia2018}. The coupling to the Maxwell theory in four dimensions has been recently studied using a Kaluza-Klein approach  from Ho\v{r}ava-Lifshitz theory in five dimensions  \cite{BellorinRestucciaTello2018b,RestucciaTello2020}. 
The speed of propagation of the gravitational and electromagnetic physical modes, at low energies, is the same in agreement with the recent experimental data arising from the detection of gravitational and electromagnetic waves generated by the same source \cite{Abbott2017}. An analysis of the theory including  all higher order spatial derivative terms of the potential  was performed in \cite{BellorinRestuccia2016B}.
Although at the linearized level Ho\v{r}ava-Lifshitz theory and GR coincide, the full theories behave in a different way as can be seen from the static spherical symmetric solutions of the field equations. In this case, in the Ho\v{r}ava-Lifshitz theory there are solutions with  a throat, connecting an asymptotically flat manifold with a non-asymptotically flat one \cite{Bellorin2014},  see also \cite{Tello2021} where a charged throat is considered.
The solutions depend in one coupling constant $\alpha$ in a way that in the limit when $\alpha \to 0 $ the geometry outside the throat, on the asymtotically flat side, tends to the geometry of the Schwarzchild solutions in GR outside the black hole. The asymptotic behaviour of these solutions define the asymptotic flat behaviour in Ho\v{r}ava-Lifshitz gravity.

Non-linear interaction terms in gravity theories are essential in the propagation of the physical degrees of freedom. However, in GR in the wave zone, the propagation of the physical degrees of freedom, at leading order, is free from such non-linear terms. We may wonder if such is the case for the  Ho\v{r}ava-Lifshitz gravity, since both theories propagate exactly the same physical degrees of freedom. In particular, we are interested in the effect of the terms that break the relativistic symmetry on the propagation in the wave zone.

In \cite{mestra2021} we analyzed the wave zone in Ho\v{r}ava-Lifshitz gravity at low energies that is, when terms in the potential with high order spatial derivatives are dismissed. We found that the propagation of the physical degrees of freedom is described by the wave equation as in GR, with speed of propagation given in terms of the coupling parameter $\beta$. 

In this paper we show that there exists a wave zone on the complete Ho\v{r}ava-Lifshitz theory. We include in our analysis all high order derivative terms in the potential. We consider the theory at the kinetic conformal point, where $\lambda =1/3$. We notice that the value $\lambda = 1/3$ is protected from quantum corrections by the presence of a second class constraint which arises directly from the formulation of the theory \cite{BellorinRestucciSotomayor2013}.  

On this zone the constraints of the theory can be solved in powers of $1/r$. In distinction to GR where the Hamiltonian constraint is of first class, in this theory there are two second class constraints which allow to obtain the components $g^T$ and $N$ in terms of the transverse traceless tensorial TT modes. The lapse $N$ is not a Lagrange multiplier in this theory. 
We analyse the general coordinate transformations on the wave zone and obtain the field equations for the physical degrees of freedom described by the TT modes of the metric.
At order $1/r$, the field equations become a linear partial differential equation, which reduces to the wave equation at low energies.
Although this field equation at order $1/r$ is the same that arises form the linearized theory, the nontrivial solution for the $g^T$ and $N$ fields provides the existence of a Newtonian background absent in the linearized theory and relevant in the asymptotic behaviour of the theory, in particular in the determination of the gravitational energy. 
In the formulation of the theory on the wave zone, we follow the approach rigorously established in \cite{Arnowitt1961}.

In distinction to the ADM analysis of GR the solution of the constraints and the dynamical equations involve an elliptic operator of sixth order, yielding a higher order dispersion relation. 
The terms which break the Lorentz symmetry in the action of the theory contribute to the Newtonian background and appear explicitly in the expression of the gravitational energy. In order to have a finite gravitational energy some restrictions have to be imposed on the coupling parameters of the potential. In fact, the solution of the constraints involve an evolution operator with high order in spatial derivatives, which may have zero modes. They  produce inconsistent contributions to the gravitational energy. In order to avoid them we impose restrictions to the coupling parameters.

The relevance of the existence of a wave zone in a theory describing gravity is directly related to the detection of gravitational waves (GW) in the last five years.  
In 2010, they were measured indirectly through the relative reduction of the distance between the members of a binary pulsar system \textbf{PSR B1913+16} by continuos emission of GW \cite{Weisberg_2010}. In 2015, it was detected directly by LIGO-Virgo collaboration the first signal of a GW produced from two black hole-black hole  (BHBH) coalescence \textbf{GW150914}  \cite{Abbott2016,Abbott2016A,Abbott2018a}. After that, there have been  multiple detections of GW due to coalescence of compact-binary system as BHBH, neutron start-neutron start (NSNS) and neutron star-black hole (NSBH) \cite{Miller2019,Abbott2019,abbott2021observation,abbott2021gwtc,nitz20213}. 

 In the case of \textbf{GW170817} event, also it was detected its electromagnetic signal, $\gamma$ rays burst \textbf{\,\,\,GBR170817A}  \cite{Abbott_2017GRB}, which has been very important to compare the speeds of both electromagnetic and gravitational signals, it constrains the difference between the speed of gravity and the speed of light to be between $-3\times {10}^{-15}$ and $+7\times {10}^{-16}$ times the speed of light \cite{Abbott_2017,Abbott2019b}.
In the Ho\v{r}ava-Lifshitz at the kinetic conformal point theory both interactions have the same speed of propagation \cite{BellorinRestucciaTello2018b, RestucciaTello2020}.

A main difference between the wave zone in Ho\v{r}ava-Lifshitz theory and that in GR is the modified dispersion relation (MDR) involving a polynomial in the square of the wave number $k$, which at low energy and for $\beta=1$ reduces to the relativistic relation. 
Modifications of Dispersion Relations (MDR) in theories with Lorentz Invariance Violations (LIV) have been discussed in the literature, for a review of experimental test of LIV in astroparticle physics scenarios see \cite{Martinez-Huerta-2020} and its references.
The Large Hight Altitude Air 
Shower Observatory-(LHAASO) have recently  detected  Ultra-Hight-Energy-photons  (UHE $\gamma$-rays) in the PeV energy scale \cite{cao2021ultrahigh}. These observations suggest the existence of subluminal LIV beyond Special Relativity in the photon sector at the scale of $3,6\times 10^{17}$GeV \cite{li2021ultrahigh}. In concordance with  references \cite{xu2016(1),xu2016(2),xu2018} in which it is stated that, from data  of GRBs observed by the Fermi Gamma-ray Space Telescope (FGST),  to  this same  energy-scale the speed of light is energy dependent.

\section{Foliations in Ho\v{r}ava gravity}
We consider a 3-dimensional foliation of a 4-dimensional manifold $M$. It is a decomposition of $M$ into a union of disjoint 3-dimensional submanifolds, the leaves of the foliation, such that the covering of $M$ by charts $U_i$ together with homeomorphisms  $\varphi_i : U_i \iff \mathcal{N}_i \subset \mathbb{R}^{4} $ satisfy for overlapping pairs $U_i, U_j$:
\begin{equation}
	\varphi_j \circ \varphi_i^{-1}  : \mathbb{R}^{4} \to  \mathbb{R}^{4} , 
	\end{equation}
	is a $C^\infty$ bijection from $\varphi_i (U_i \cap U_j)$ onto $\varphi_j(U_i \cap U_j)$ and %
	\begin{equation}
	\label{bijection}
	\varphi_j \circ \varphi_i^{-1}  : (t,x) \to (\tilde{t}(t), \tilde{x}(t,x)) \,,
	\end{equation}
	where $x$ are coordinates on the 3-dimensional leaves, $\Sigma_{t}$, and $t$ on the codimension $1$ manifold which we take to be $R$. We assume that the leaves are Riemannian manifolds with metric $g_{ij}$. In the Ho\v{r}ava-Lifshitz formulation $t$ and $x^i$, $i=1,2,3$ have different dimensions and there is no a priori universal constant of dimension $[T]^{-1}[L]$. Although there are couplings with this dimension they run with the energy scale, hence only on the infrared limit (or UV one) we can have such constant. 
	Although there is not a metric on the $M$ manifold, there are intrinsic geometrical objects which we define: 
	\begin{equation}
	  N(t,x) \, dt,  
	\end{equation}
	with dimension $[T]$ and 
\begin{equation}
	  dx^{i} + N^i(t,x) dt,  
	\end{equation}
with dimension $[L]$. They transform under (\ref{bijection}) :
\begin{eqnarray}
\label{transformationN}
\tilde{N}(\tilde{t}, \tilde{x}) d \tilde{t} = N(x,t) dt,  \\
\label{transformationNi}
d\tilde{x}^{i} + \tilde{N}^{i}(\tilde{t}, \tilde{x}) d \tilde{t}= \frac{\partial \tilde{x}^{i}}{\partial x^{j}}[dx^{j} + N^{j}(t,x) dt],
\end{eqnarray}	
and allow to have an intrinsic volume element

\begin{eqnarray}
N dt \wedge (dx^{1} + N^{1}dt) \wedge (dx^{2} + N^{2}dt) \wedge (dx^{3} + N^{3}dt) \sqrt{g}= \nonumber \\  
N dt \wedge dx^{1} \wedge dx^{2} \wedge dx^{3} \sqrt{g}\, ,\qquad
\end{eqnarray}
where $g$ is the determinant of the 3-dimensional metric $g_{ij}$ on the leaves. The metric is taken to be a dimensionless tensor under the diffeomorphisms on the leaves.

From (\ref{transformationN},\ref{transformationNi}) we obtain the transformation law of $N$, a dimensionless density, and $N^i$. In the Ho\v{r}ava-Lifshitz formulation of gravity $t$ scales as $b^z$ while $x^i$ as $b^1$,
consequently, the dimension of $t$ is $[t]=[L]^z$  and that of $N^i$ is $[N^i]=[L]^{1-z}$. We notice that the  above construction is generic, for any value  of $z$.
In the 4-dimensional Ho\v{r}ava-Lifshitz gravitational theories $z=3$ in order to have a dimensionless overall coupling constant. Hence $[N^i]=[L]^{-2}$ and as we have determined $[N]=[L]^0$.

The transformation law for $N$ and $N^i$ under 
\begin{equation}
\label{diff}
\tilde{t}=\tilde{t} (t) \quad ,  \qquad \tilde{x}^{i}=\tilde{x}^{i}(t,x),
\end{equation}
is then
\begin{eqnarray}
\label{transformation2}
\tilde{N}(\tilde{t}, \tilde{x}) \dot{\tilde{t}}(t) = N(t,x) \,, \\
\label{transformation3}
\tilde{N}^{i} (\tilde{t}, \tilde{x}) \dot{\tilde{t}}(t)=-\dot{\tilde{x}}^{i}(t,x) + \frac{\partial \tilde{x}^{i}}{\partial x^{j}}N^{j}(t,x),
\end{eqnarray}
respectively, $()\dot{}\equiv \frac{\partial}{\partial t}()$. We notice that $N$ and $N^i$ transform as densities under time reparametrization; $N^i$ does not transform as a vector field under diffeomorphisms on the leaves. The first term on the right hand side member of (\ref{transformation3}) is characteristic of the transformation law of a Lagrange multiplier. 
If we rewrite the general finite transformation of coordinates as
\begin{equation}
\label{coord_general_transformation}
\tilde{t} = t + f(t) \, , \qquad    \tilde{x}^{i} = x^{i} + \xi^{i} (t,x),
\end{equation}
$f$ and $\xi^{i}$ are  $C^{1}$ arbitrary functions, we obtain
\begin{equation}
\label{transformation_N}
\tilde{N}(\tilde{t},\tilde{x})(1+\dot{f}(t)) = N(t,x) \,, 
\end{equation}
\begin{equation}
\label{transformation_Ni}
\tilde{N}^{i} (\tilde{t},\tilde{x})(1+\dot{f}(t)) = -\dot{\xi}^{i}(t,x) + N^{i}(t,x) + \frac{\partial \xi^{i}}{\partial x^{j}}N^{j} (t,x) \,,
\end{equation}
and also
\begin{equation}
\label{transformation_g}
	g_{i j}(t,x) = \frac{\partial \tilde{x}^{l}}{\partial x^{i}} \frac{\partial \tilde{x}^{m}}{\partial x^{j}}\tilde{g}_{lm}(\tilde{t},\tilde{x}).
	\end{equation}	
\subsection{The T+L decomposition}
In the following analysis we will use the T+L decomposition of symmetric tensors fields ($f_{ij}$) vanishing at infinity in terms of linear orthogonal symmetric parts \cite{ArnowittDesserMisner2008},
\begin{equation}
\label{T+L}
f_{ij}=f_{ij}^{TT}+f_{ij}^{T}+2f_{(i,j)} \,.
\end{equation}
	The transverse part $f_{ij}^{T}\equiv\frac{1}{2}[\delta_{ij}f^{T}-\frac{1}{\Delta}f_{,ij}^{T}]$
	satisfies  $f_{ij,j}^{T}=0$. The transverse trace-less  $f_{ij}^{TT}$
	satisfies  $f_{ij,j}^{TT}=0$ and $f_{ii}^{TT}=0$.   The remaining term  $2f_{(i,j)}$ is its longitudinal part. $f^{T}=f_{ii}^{T}$ is the trace of the transverse part of $f_{ij}$.   $\frac{1}{\Delta}$ is the inverse of the flat space Laplacian,  defined on the space of functions which vanishes at infinity.
	
	The tensor components are

	\begin{eqnarray}
	f_{i,j}&=&\frac{1}{\Delta} \left\{f_{ik,kj}-\frac{1}{2}\left[\frac{1}{\Delta}f_{lm,lm}\right]_{,ij}\right\},\\
	f_{ij}^{T}&=&\frac{1}{2}\left[f^{T}\delta_{ij}- \frac{1}{\Delta}f_{,ij}^{T}\right],\\
	f^{T}&=&\frac{1}{\Delta}\left(f_{ll,mm}-f_{lm,lm}\right),\\
	f_{ij}^{TT}&=&f_{ij}-f_{ij}^{T}- 2f_{(i,j)} \,.
	\end{eqnarray}

\section{The 3+1-Ho\v{r}ava-Lifshitz gravity at the conformal kinetic point}
	The 3+1 anisotropic Ho\v{r}ava-Lifshitz Hamiltonian at the conformal kinetic point is given by 

\begin{eqnarray}
    H=\int_{\Sigma_{t}} d^{3}x\bigg\{N\sqrt{g}  \left[ 
	\frac{\pi^{ij}\pi_{ij}}{g}- \mathcal{V}\left(g_{ij},N\right)
	\right]  \nonumber \\
	- N _{j}H^{j}- \sigma P_{N}-\mu \pi \bigg\}+\beta E_{ADM}.\label{Hamiltonian}
\end{eqnarray}
Where  $( N_{i},\sigma,\mu)$
	are Lagrange multiplier,  $(H^{j}, P_{N},\pi)$ are primary constraints, and  the  potential $\mathcal{V}= \mathcal{V}^{(1)}+\mathcal{V}^{(2)}+\mathcal{V}^{(3)}$ is the most general scalar constructed from the 3-dimensional spacelike metric and the  lapse $N$-function. It is independent of the conjugate momenta and of the  $N_i\equiv g_{ij}N^{j}$. The expression of the potential, where only the interacting terms which will contribute to the wave zone are explicitly given, is the following

	\begin{eqnarray}
	\label{potential1}
	\mathcal{V}^{(1)}&=& \beta R + \alpha a_ia^i, \\
	\label{potential2}
	\mathcal{V}^{(2)}&=& \alpha_{1} R \nabla_i a^i +  \alpha_{2} \nabla_i a_j \nabla^i a^j  +  \beta_{1} R_{ij} R^{ij} \nonumber \\&&+ \beta_{2} R^{2} + O^{(2)}(a_ia^i),\\
	\label{potential3}
	\mathcal{V}^{(3)}&=& \alpha_{3}\nabla^2 R \nabla_i a^i +  \alpha_{4} \nabla^2 a_i \nabla^2 a^i +  \beta_{3} \nabla_i R_{jk}\nabla^i R^{jk} \nonumber \\&&+ \beta_{4} \nabla_i R \nabla^i R + O^{(3)}(R_{ij}, a_k) \,,
	\end{eqnarray}
here, 	$a_i\equiv \frac{1}{N}\partial_{i} N$,    is a 3-vector under transformations (\ref{diff}) with dimension $[L]^{-1}$  \cite{Blas2010},
	and  $\alpha$'s and $\beta$'s are coupling constants. We notice that couplings involved in (\ref{potential3}) are dimensionless. The terms $O^{(2)}$ and $O^{(3)}$ contain products of more than two fields.   They will not contribute to the  leading order term neither to the next order one in the wave zone.

The  surface integral, 
\begin{equation}
\label{ADMenergy}
  E_{ADM}\equiv \oint_{\partial \Sigma_{t}} \left( \partial_{j}g_{ij}-\partial_{i}g_{jj}\right)dS_{i},\,
\end{equation}
is added in order to ensure the differentiability of the Hamiltonian with respect to the metric, see \cite{Regge-Teitelboim1974} where this idea was introduced for RG. $E_{ADM}$ is the well known ADM-energy in General Relativity.

	The primary constraints,
	\begin{eqnarray}
	\label{P-constraint}
	\pi\equiv g_{ij}\pi^{ij}=0,\\
	\label{P_N-constraint}
	P_{N}=0,
	\end{eqnarray}
	are  second class, while the momentum constraints,
	\begin{eqnarray}
	\label{Pi-transverse-cosntraint}
	H^{j} \equiv 2 \nabla_{i} \pi^{ij}=0,
	\end{eqnarray}
	are  first class ones.
	
	The conservation of the primary constrains  (\ref{P-constraint}, \ref{P_N-constraint}) yields the constraints 
	\begin{eqnarray}
	\label{H_P-constraint}
	H_{P} &\equiv& \frac{3}{2} \frac{N}{\sqrt{g}} \pi^{ij}\pi_{ij}+\frac{1}{2}N\sqrt{g}\beta R
	+ N\sqrt{g}\left(\frac{\alpha}{2}-2\beta \right)a_{i}a^{i}\nonumber \\ 
	&&- 2\beta N\sqrt{g}\nabla^{i}a_{i} + \{\pi, \mathcal{U}\}_{PB}=0,\\
	\label{H_N-constraint}
	H_{N}&\equiv& -\frac{1}{\sqrt{g}}\left(\pi^{ij}\pi_{ij}- \beta g R \right)- \alpha \sqrt{g}a_{i}a^{i} \nonumber \\ &&-2\alpha \sqrt{g}\nabla_{i}a^{i} + \{P_N, \mathcal{U}\}_{PB}=0 \,,
	\end{eqnarray}

	where 
	\begin{equation}
	 \mathcal{U} \equiv -\int_{\Sigma_{t}} d^{3}x N \sqrt{g}(\mathcal{V}^{(2)}+\mathcal{V}^{(3)}) \,.
	\end{equation}
	
	These are  second class constraints. Conservation of  (\ref{H_P-constraint}, \ref{H_N-constraint}) determine Lagrange multipliers. Then, (\ref{P-constraint}), (\ref{P_N-constraint}), (\ref{Pi-transverse-cosntraint}), (\ref{H_P-constraint}) and (\ref{H_N-constraint}) is the complete set of constraints.
	
	The dynamical field equations are 
	\begin{eqnarray}
	\label{gpunto}
	\dot{g}_{ij}&=&\frac{2N}{\sqrt{g}}\pi_{ij}+ 2\nabla_{(i}N_{j)}- \mu g_{ij}, \\
	\label{pipunto}
	\dot{\pi}^{ij}&=&\frac{N}{2}\frac{g^{ij}}{\sqrt{g}}\pi^{kl}\pi_{kl}-  \frac{2N}{\sqrt{g}}\pi^{ik}\pi^{j}{}_{k}+N \sqrt{g}\beta \left(\frac{R}{2} g^{ij}-R^{ij}\right) \nonumber \\
	&&\nonumber \\ &&
	-\alpha N\sqrt{g}\left(a^{i}a^{j}-\frac{1}{2}g^{ij}a_{k}a^{k}\right)- \nabla_{k}\left[2\pi^{k(i} N^{j)}-\pi^{ij} N^{k}\right]\nonumber \\
	&&+\beta\sqrt{g}\left[\nabla^{(i}\nabla^{j)}N-g^{ij}\nabla^{2}N\right]+\mu \pi^{ij} + \{\pi^{ij}, \mathcal{U}\}_{PB} \,,
	\end{eqnarray}
where the last term in (\ref{pipunto}) will  contribute with high order in spatial derivatives terms to the wave zone.

The Hamiltonian (\ref{Hamiltonian}) can be rewritten in terms of the constraints of the theory, up to total divergence. In fact, (\ref{H_N-constraint})  can be written as
\begin{eqnarray}
	H_{N}\equiv\sqrt{g}\left( \frac{\pi^{ij}\pi_{ij}}{g} -\mathcal{V}\right)-\frac{\sqrt{g}}{N}\nabla_{i}\left[-N\frac{\partial \mathcal{V}} {\partial a_{i}}\right. \nonumber\\
	\hspace{0mm} \left. +\nabla_{j}\left( N \frac{\partial \mathcal{V}}{\partial \nabla_{j}a_{i}}\right)- \nabla_{k}\nabla_{j}\left(N \frac{\partial\mathcal{V}}{\partial\nabla_{j}\nabla_{k}a_{i}}\right)+\cdots \right]=0, \qquad
	\end{eqnarray}
	where the first term arise from variation of the factor $N$ while the second terms comes from the variation of the potential. It follows then, that 
	\begin{eqnarray}
	N\sqrt{g}\left( \frac{\pi^{ij}\pi_{ij}}{g} -\mathcal{V}\right)=N H_{N}+\sqrt{g}\nabla_{i}\left[-N\frac{\partial \mathcal{V}}{\partial a_{i}} \right. \nonumber \\
	\left. + \nabla_{j}\left( N \frac{\partial\mathcal{V}}{\partial \nabla_{j} a_{i}}\right)-\nabla_{k}\nabla_{j}\left(N \frac{\partial\mathcal{V}}{\partial\nabla_{j}\nabla_{k}a_{i}}\right)+ \, \cdots \right].\qquad
	\end{eqnarray}
	Under asymptotically flat conditions: $g_{ij}=\delta_{ij}+\mathcal{O}\left( \frac{1}{r}\right)$, $N=1+\mathcal{O}\left( \frac{1}{r}\right)$, $N_{i}=\mathcal{O}\left( \frac{1}{r}\right)$, $\pi^{ij}=\mathcal{O}\left( \frac{1}{r}\right)$, and all first spatial derivatives of fields  of order $\mathcal{O}\left( \frac{1}{r^{2}}\right)$, the total divergence that contributes at infinity arises from the first term in the parenthesis. It is
	\begin{equation}-2\alpha\sqrt{g}\nabla_{i}\left(g^{ij}\nabla_{j}N\right).
	\end{equation}
	We may now evaluate the physical Hamiltonian density in a particular coordinate system. We proceed as in \cite{mestra2021}. The Lagrangian evaluated on the constraints submanifold is given by
	\begin{equation}
	    L=\int dt d^{3}x\left(\pi^{ij}\dot{g}_{ij}-\mathcal{H} \right)- E_{ADM},
	\end{equation}
	where $\mathcal{H}\equiv -2\alpha \sqrt{g}\nabla_{i}\nabla^{i}N+ \text{total divergence}$. The generic terms "total divergence"  do not contribute to the Lagrangian under the asymptotically flat conditions. The Lagrangian $L$ can be expressed in terms of the T+L components as in \cite{mestra2021}. We may consider the coordinate condition 
	\begin{equation}
	\label{coordinate_election}
	 g_{i}=x^{i}+ \left(\frac{1}{4\Delta} \right) g_{,i}^{T}\, , \quad \Delta\equiv \delta^{ij}\partial_{i}\partial_{j}, \end{equation}
	 where $g_{i}$ is the longitudinal part of the metric, and we end up, following \cite{mestra2021} with 
	 \begin{equation}
	 \mathcal{H}=-\beta \Delta g^{T}-2\alpha \sqrt{g}\nabla_{i}\left(g^{ij}\partial_{j}N\right),
	 \end{equation}
	 as the Hamiltonian density. $\Delta g^{T}$ and $\Delta N$ are obtained from the constraints (\ref{H_P-constraint},\ref{H_N-constraint}) in terms of the $g_{ij}^{TT}$, $\pi^{ijTT}$ physical components.
	 
The gravitational energy is then given by
	 \begin{equation}
	     E=\int dt\, d^{3}x\,
\mathcal{H}=\oint_{\partial \Sigma_{t}} \left( -\beta g_{,i}^{T}-2\alpha N_{,i}\right)dS_{i},	 \end{equation}
which is, of course, independent of the coordinate condition (\ref{coordinate_election}).

In order to obtain $g^{T}$ and $N$ from the constraints (\ref{H_P-constraint},\ref{H_N-constraint}) a condition on the invertibility of the higher order derivative operator must be imposed. We discuss this point in the section (\ref{Constraints resolution}). The invertibility condition, expressed as the absence of the zero modes, is assured once several conditions on the coupling parameters are imposed.
\section{The Wave zone definition }
The  "wave zone" is an asymptotic region in space within which the degrees of freedom of the theory represent free radiation.  That is, the leading order satisfies a wave equation and represents traveling spherical waves that escape to infinity  without being affected  by the source. 

In the wave zone the self-interactions do not affect the dominant order. In the analysis the non-linearities are not eliminated a priori, indeed there can exist non-trivial static terms of the same order of the leading canonical one, but nevertheless they do not affect its  propagation as free-radiation. This approach is different from the first-order perturbation analysis where non-linear effects are eliminated from the beginning. 
The existence of a wave zone  is a very important property of non-linear theories describing long range interactions.

In linear theories like classical electrodynamics, non-linearities may be due to sources and they do not occur if we take distances far enough from them. Then in this case, if radiation has a wave-number $k$, the wave zone coincide with the "far zone" $kr\gg1$. 
That is, within spatial distance between source and the wave zone, $r$, there fit a large number of wavelengths and it guarantees that  gradients and temporal derivatives of canonical modes  are $\mathcal{O}(1/r)$.
In general, we can apply the time-Fourier-transformation to the wave equation, it becomes a Helmholtz equation  $\Delta u+ k^{2}u=0$ and one can write its solution in the form $u=\frac{e^{ikr}}{r} \sum_{n=0}\frac{f_{n}(\theta,\phi)}{r^{n}}$.
Then it is straightforward to prove that  if the $f_0$-coefficient of the $1/r$- term  is zero then $u=0$, that is, in the wave zone the only non-zero solution of the wave equation contains always the $\mathcal{O}(1/r)$ contribution. 
This is different from the solution of the Laplace equation. In fact, even if the monopole, dipole, etc, contributions are zero, the multipolar contributions can be nonzero.

In general, for non-linear theories the propagating fields may  act as effective sources of itself. This self-interaction phenomena break-up the desired free-radiation behavior. \,\, Therefore the far zone condition is not sufficient to define the wave zone. The additional requirements  have to guarantee  that the field amplitudes   are small such that   the non-linear terms are "negligible". Here "negligible" means that it does not affect the free-wave propagation.

For GR there exists a well-defined  wave zone, in which the background-curvature does not affect the radiation of canonical modes. The gauge invariant fields have the following behaviour:  The oscillatory part of  $g_{ij}^{TT}\sim \pi^{ijTT}=\mathcal{O}(1/r)$, the oscillatory part of $g_{ij}^{T} \sim \mathcal{O}(1/r^{2})$, the static part of  $g_{ij}^{T} =\mathcal{O}(1/r)$ and of $\pi^{ijL}=\mathcal{O}(1/r^{2})$.
The remaining  parts of $g_{ij}$ and $\pi^{ij}$ and   ADM variables $(N, N^{i})$ are gauge dependent and they do not affect the propagation of canonical modes \cite{Arnowitt1961}.

The wave zone split the space into three regions: the interior,  wave zone and the exterior. The "interior" region bounded by the sphere of radius $R$ centered at the origin,  $B_{R}(0)$,  contains the sources and possibly space-time singularities; here the phenomena of self-interaction are not negligible even at points where sources are not located. 
The wave zone, $B_{R_{1}}(0)/B_{R}(0)$ with $ R<r<R_{1}$, where the curvature does not affect radiation. Finally the "exterior" region,  $\mathbb{R}^{3}/B_{R_{1}}(0)$, where the propagating fields decay rapidly. This represents an asymptotically flat space-time.  

In this work  we prove that in the 3+1 non-projectable Ho\v{r}ava-Lifshitz theory at the kinetic-conformal point,  there exists a wave zone as it does in GR. 
That is, the  $(g_{ij}^{TT},\pi^{ijTT})\sim f(\theta, \phi)\exp i(kx-\omega(k)t)/r$, with $f(\theta, \phi)=\mathcal{O}(1)$, propagate without being affected by the static parts in the same way as it happens in the linearized theory.  
Due to the anisotropic character of this theory the dispersion relation contains, besides the terms $\omega^{2}$ and $k^{2}$, $k^{4}$ and $k^{6}$  terms associate to higher order derivative terms, this represents a propagating dispersive signal. 
At low energies the dispersion reduces to the usual non-dispersive wave and differs from GR in the speed of propagation. The speed of propagation at low energies is equal to $\sqrt{\beta}$ and the recent gravitational waves observations fixed it to $1$ the GR value, up to an error of the order of one part in $10^ {15}$\cite{Abbott_2017,Abbott2019b,EmirGumrukcuoglu2018, ramos2019}.  

We define the wave zone in Ho\v{r}ava-Lifshitz  theory, as in GR, to be the space like domain far away from the sources satisfying the conditions:
\begin{enumerate}
		\item[$i)$]   
		$kr\gg1$. $k$ is the wave number, $r=(x_{i}x^{i})^{1/2}$. 
		\item[$ii)$]
		
		\begin{eqnarray}
        \label{wave zone metric}|g_{ij}-\delta_{ij}| \lesssim A/r, \\
        \label{wave zone N}\quad |N-1|\lesssim A/r, \\ \label{wave zone Ni}\quad |N^{i}|\lesssim A/r ,
		\end{eqnarray}

where  $A(t,\theta,\phi)$  represents  functions of time and angles such that $A$ and all its derivatives are bounded.  
		\item[$iii)$] 
	
		\begin{eqnarray}|\partial g_{ij}/\partial(kr)|^{2}\ll|g_{ij}-\delta_{ij}|\,, \\ |\partial N_{i}/\partial(kr)|^{2}\ll|N_{i}|\,, \\ |\partial N/\partial(kr)|^{2}\ll|N-1|\,, 
		\end{eqnarray}
	
		\end{enumerate}  
 are fulfilled. Here we used the short notation \,\, $\lesssim A/r$  $\equiv$  $\mathcal{O}(A/r)$, thus  "$\lesssim A/r$" means the  left part decrease at least as $A/r$. And $(g,\delta)$ are symbolic representations of  any component of the metric and the euclidean tridimensional metric tensor. 

	The three conditions can be satisfied by choosing a region with large enough distance from the source. The first condition is the same one as in linear theories and indicates that the wave zone is quite far from the sources (in distance measured in units of wavelength) and guarantees that the  gradients of the canonical modes are $\mathcal{O}(1/r)$ and exclude the  wave number  $k\rightarrow 0 $ in the wave zone.
	The second and third conditions are necessary to ensure that, in the wave zone, there are no self-interactions and the canonical variables propagate freely. The second condition is imposed in order to ensure that not only the perturbations of the dynamic modes are  $\mathcal{O}(1/r)$ but  that the Newtonian parts are also $\mathcal{O}(1/r)$, and guarantees that the terms of higher order than the leading one in $|g_{ij}-\delta_{ij}|$ remain negligible compared to the leading order. 
	The third condition is taken in order to ensure that the non-linear terms containing spatial derivatives are small compared to the leading order. 
	This implies that for a wave-number $k>k_{min}$  the radiation can be treated as free radiation if $k_{min}\gg k_{max}(A/r)^{1/2}$,  here $|g-\delta|\sim A/r$ and $k_{max}$ is the maximum frequency. It ensures that the interference of two or more sub-leading $\mathcal{O}(1/r^{2})$ modes do not  generate an effect comparable to leading $\mathcal{O}(1/r)$.

Conditions  (\ref{wave zone metric},\ref{wave zone N},\ref{wave zone Ni})  are  preserved under a change of coordinates (\ref{diff}) if and only if 
\begin{eqnarray}
	|\dot{\xi}^{i}|\lesssim \frac{A^{i}}{r}  \ll 1,\quad
|{\xi^i}_{,j}|\lesssim \frac{{A^{i}}_j}{r} \ll 1,\quad
	|\dot{f}(t)| \lesssim \frac{A}{r}  \ll 1. \nonumber\\
	\end{eqnarray}	
 We then have, besides (\ref{transformation_N},\ref{transformation_Ni},\ref{transformation_g}), 

\begin{equation}
    \tilde{N}(1+\dot{f})=N,
\end{equation}
\begin{equation}
\label{transformation_Ni_estimated}
\tilde{N}^{i} (\tilde{t},\tilde{x})(1+\dot{f}(t)) = N^{i}(t,x) - \dot{\xi}^{i}(t,x) +  \mathcal{O} (1/r^{2})\,,
\end{equation}
\begin{equation}
\label{transformation_g_estimated}
\tilde{g}_{ij}(\tilde{t},\tilde{x}) - g_{ij}(t,x) = - (\xi_{i,j}(x,t) + \xi_{j,i}(x,t) ) +\mathcal{O} (1/r^{2}).
\end{equation}

We remark that $\frac{\mathcal{O}(1/{r^2})}{1/{r}} \ll 1$. 

We then conclude that the longitudinal part of $g_{ij}$ is a gauge dependent field. It can be fixed by choosing particular coordinates on the leaves.
We notice that the admissible coordinate transformations are not necessarily infinitesimal ones. In fact, one may have a parameter $\xi \sim  \log r$. Besides, one may have parameters with dependence $1/r$ and also $e^{ikr}/r$ or $\mathcal{O} (1/r^{2})$. Consequently, the continuous transformations on the wave-zone are an extension of the infinitesimal ones. These ones have the usual form
	\begin{eqnarray}
	\tilde{N}({t},{x}) - N(t,x) &=& -\xi^{k} \partial_k N (t,x) - \dot{f}(t)N(t,x) \nonumber \\
	&&- f(t)\dot{N}(t,x)\,,\\
	\tilde{N}^{i}({t},{x}) -N^{i}({t},{x}) &=&  -\xi^{k} \partial_k N^{i}({t},{x}) + \partial_j \xi^{i}N^{j}({t},{x})-\dot{\xi}^{i} \nonumber \\ &&- \dot{f}(t)N^{i}(t,x) - f(t)\dot{N}^{i}(t,x) \,, \\
	\tilde{g}_{ij}({t},{x}) - {g}_{ij}({t},{x}) &=&- \xi^{k} \partial_k g_{ij}({t},{x})-\nonumber \\ &&g_{ki} \partial_j \xi^k - g_{kj} \partial_i \xi^k - f\dot{g}_{ij}({t},{x}) \,, \qquad
	\end{eqnarray}
	where in this case, $f(t)$ and $\xi^i (t,x)$ are infinitesimal parameters. We notice that these transformations contain terms which are of order $\mathcal{O} (1/r^2)$, hence in the wave-zone at order $\mathcal{O}(1/r)$ they can be eliminated.
	
\section{Behavior in the wave zone}
In the wave zone the metric, lapse and shift fields have the  behavior 
	\begin{eqnarray}
	\label{asymptotic-metric-deviation}
	g_{i j}-\delta_{i j}\sim N_{i}\sim N-1 &\lesssim& \frac{B_{ij}}{r}+\frac{A_{ij} e^{ikr}}{r},
	\end{eqnarray} 
	\begin{eqnarray}
	\label{asymptotic-spacial-temporal-metric-spatial-derivative}
	N_{i,j}\sim g_{ij,l} \sim\Gamma^{i}_{\hspace{0.1cm}jl} \lesssim \frac{B}{r^{2}}+k\frac{A e^{ikr}}{r}\,,
	\end{eqnarray}
	where $A$ and $B$ are a generic functions of time and angles  with bounded derivatives.

	\subsection{ Laplacian's solutions in the wave zone}
	Let $\varphi \in C^{2}(\Omega)$ be a   solution of  Poisson equation $\Delta\varphi=-4\pi \rho$ in $\Omega\equiv \mathbb{R}^{3}/B_R(0)$   such that  $\varphi \rightarrow 0$ when $r\rightarrow \infty$. 
	Expanding $\varphi(r,\theta,\phi)$ into spherical harmonics $Y_{lm}(\theta,\phi)$ a standard result gives, 
	\begin{eqnarray}
	\label{phi-expansion}
	\varphi(r,\theta,\phi)&=& \sum_{l=0}^{l=\infty}\sum_{m=-l}^{m=l} \chi_{lm}(r)r^{l}Y_{lm}(\theta,\phi),\\
	\label{chi_lm}
	\chi_{lm}(r)&\equiv& \int_{r}^{\infty}\frac{M_{lm}(r')}{r'^{2l+2}}dr',\\
	\label{M_lm}
	M_{lm}(r)&\equiv&\int_{B_{r}(0)} \rho(r',\theta',\phi')r'^{l}Y'_{lm}(r',\theta',\phi') d^{3}r'.\quad
	\end{eqnarray}
	Following Arnowitt, Desser and Misner \cite{Arnowitt1961} in the wave zone we have:
	
	\begin{itemize}
		
		\item[$i$)] If in the region $r>R$ the source $\rho$ has a  oscillatory-asymptotic-behavior $\rho\sim Y_{lm} e^{ikr}/r^{n}$   then,

		\begin{equation}
		\label{Oscilatory-asymptotic-behavior}
		\varphi\sim \left(1/k^{2}\right) Y_{lm}\frac{e^{ikr}}{r^{n}} \left[1+\mathcal{O}(1/kr)\right]+ \sum_{l=0}^{l=\infty}\sum_{m=-l}^{m=l}c_{lm} Y_{lm}\frac{1}{r^{l+1}}.
		\end{equation} 

		Here the  $c_{lm}$ do not depend on the  spatial coordinates and they are   determined from the behavior of $\rho$ in the  region $r\leq R$.  Hence, an oscillatory source produce in $r>R$  a  source-independent oscillatory solution of the same order of $\rho$ plus a static source-depend part $\lesssim 1/r$. By static we mean non-oscillatory on the space dependence.

		\item[$ii$)] If in the region $r > R $, 
		\begin{equation}\psi\sim B_{m}(t,\theta,\phi)/r^{m}+ A_{n}(t,\theta,\phi)e^{ikr}/r^{n},\end{equation}
		with   $|B_{m}(t,\theta,\phi)|$ and $|A_{n}(t,\theta,\phi)|$ bounded, then

		\label{inversionlaplacianoconderivadas}
		\begin{eqnarray}
		\label{first-derivative-source}
		\frac{1}{\Delta}\psi,_{j}&\sim&\frac{\langle B \rangle}{ r^{m-1}} + \frac{1}{ik}\frac{\langle A \rangle  e^{ikr}}{ r^{n}}\nonumber \\
		&&+ \sum_{l=1}^{\infty}\sum_{p=-l}^{p=l}c_{lp}Y_{lp}\frac{1}{ r^{l+1}}, \quad (m \geq 2).\\
		\label{second-derivative-source}
		\frac{1}{\Delta}\psi,_{ij}&\sim&\frac{\langle B \rangle}{r^{m}} + \frac{\langle A \rangle  e^{ikr}}{r^{n}}\nonumber \\ 
		&&+\sum_{l=2}^{\infty}\sum_{p=-l}^{p=l}c_{lp} Y_{lp}\frac{1}{r^{l+1}},\quad (m \geq 1).
		\end{eqnarray}
		
		$\langle A \rangle$, etc, mean generic angular integrals of $A$ and other angular functions. For $m\geq 3 $ the expressions (\ref{first-derivative-source}) and (\ref{second-derivative-source}) may also have contributions of the form  $r^{-m+1}\log r$ and $r^{-m}\log r$ respectively, which are source independent.
		
	\end{itemize}
		
		\section{The wave zone in Ho\v{r}ava-Lifshitz gravity at the kinetic conformal point}

	We now analyze the solution of the field equations in the wave zone and show that the equations of the  canonical transverse-traceless modes describe free propagating fields.
	\subsection{Solution of the primary constraints } 
	Taking a derivative with respect to the $j$-coordinate of the T+L decomposition of the momentum constraint in  (\ref{Pi-transverse-cosntraint}) we obtain
	\begin{equation}
	\label{Lapalcian_pi-i,i}
	\Delta \pi^{j}_{\hspace{0.1cm},j}=-\frac{1}{2}\left(\pi^{lm}\Gamma^{j}_{\hspace{0.1cm}lm}\right)_{,j}.
	\end{equation}
	If we use the estimating  of previous section, the result (\ref{first-derivative-source}), and the same argument  of finite momenta used in the wave zone for RG (see the  appendix C of \cite{Arnowitt1961})), then
	\begin{equation}
	\label{TRUE-asymptotic-divergence-PI-behavior}
	\pi^{j}_{\hspace{0.1cm},j}\lesssim\frac{B}{r^{2}}+k\frac{\tilde{A}  e^{ikr}}{r^{2}} \,,
	\end{equation}
	where $B$ is source dependent while $\tilde{A}$ is source independent.
	Taking now a derivative with respect to the $i$ coordinate of the T+L decomposition of the momentum constraint (\ref{Pi-transverse-cosntraint}) we get
	\begin{equation}
	\Delta \pi^{j}_{\hspace{0.1cm},i}=	\left(\pi^{l}_{\hspace{0.1cm},l}\right)_{,ij}-\left(\pi^{lm}\Gamma^{j}_{\hspace{0.1cm}lm}\right)_{,i} \,.
	\end{equation}
	
	Thus, we can use  (\ref{first-derivative-source},\ref{second-derivative-source}) to finally obtain 
	\begin{equation}
	\pi^{j}_{\hspace{0.1cm},i}\lesssim\frac{B_{i}^{j}}{r^{2}}+k\frac{\hat{A}_{i}^{j}  e^{ikr}}{r^{2}} \,, 
	\end{equation}
	for new functions  $B_i^j$ and $\tilde{A}_i^j$.

	The constraint  (\ref{P-constraint}) expressed on the T+L decomposition becomes $\pi^{T}+2\pi^{i}_{\hspace{0.1cm},i}=0$, then 
	\begin{eqnarray}
	\pi^{T}\lesssim\frac{B}{r^{2}}+k\frac{\hat{A}  e^{ikr}}{r^{2}},\quad
	\pi^{ijT}\lesssim\frac{B^{ij}}{r^{2}}+k\frac{\hat{A}^{ij}  e^{ikr}}{r^{2}},
	\end{eqnarray} 
	where we have used the estimate given by (\ref{second-derivative-source}).
	
	We note that  the TT term of the T+L decomposition is the dominant one among the terms with an oscillatory behavior,
	\begin{equation}
	\pi^{ijTT}\lesssim\frac{B^{ij}}{r}+k\frac{\hat{A}^{ij}  e^{ikr}}{r}.
	\end{equation}
	In the following section we will prove that the static parts of $\pi^{ij}\lesssim B^{ij}/r^{2}$, where $B^{ij}$ does not depend on time.

	\subsection{Solution of the secondary constraints } 
	Now we calculate the constraints $H_{N} = 0$ and $H_{P} =  0$. We include the $\mathcal{V}$-potential terms up to $\mathcal{O}(1/r)$. That is the relevant terms on the wave zone. 
%
%
%
We get for the terms of the form $\pi^{ij}\pi_{ij}\lesssim B/r^{2}+Ae^{ikr}/r^2$, where $B= B^{ij}B_{ij}$ .
Up to order $\mathcal{O}(1/r)$ the second class constraints $H_N=0$  and $H_{P} =0$ become
	\begin{eqnarray}
	\label{vinculo_HN}
	- \beta R + 2 \alpha \Delta N -  \alpha_{1}  \Delta R - 2 \alpha_{2} \Delta ^2N-\alpha_{3} \Delta^{2}R  \nonumber && \\ +2\alpha_{4} \Delta ^{3} N   \lesssim \frac{\tilde{B}}{r^{2}} + k^{2}P_{2}(k^{2}) \frac{\hat{A}e^{ikr}}{r^{2}}, \qquad 
	\end{eqnarray}
	
	\begin{eqnarray}
	\label{vinculo_HP}
	\frac{\beta}{2}  R - 2 \beta \Delta N - 2 \alpha_{1} \Delta^{2} N-2\alpha_{3} \Delta^{3} N \nonumber \\ - \beta_1  R^{ij}_{\,, \,ij} +\beta_{3} \Delta R^{ij}_{\,, \,ij} 
	 -  (\beta_{1}+ 4 \beta_{2})\Delta R  \nonumber \\
	+ (\beta_{3}+4\beta_{4}) \Delta^{2}R  \lesssim  \frac{\hat{B}}{r^{2}} + k^{2}P_{2}(k^{2}) \frac{\hat{A}e^{ikr}}{r^{2}}, \quad
	\end{eqnarray}
	respectively, where $\tilde{B}$ and $\hat{B}$ are proportional to $B$. We have used $P_2(k^{2})$ to denote a generic second order  polynomial on $k^{2}$ with real coefficients.
	
If we use the gauge condition $g_{ij,j}=0$, we get  
	
	 \begin{eqnarray}
	 R_{ik}+ \frac{1}{2}\Delta g_ {ik}^{TT}+ \frac{1}{2}\Delta g_ {ik}^{T}+ \frac{1}{2}g_ {,ik}^{T} \lesssim \frac{\mathcal{B}}{r^{4}} + k^{2} \frac{Ae^{ikr}}{r^{2}},\quad\\
	 R + \Delta g^{T}\lesssim \frac{\mathcal{B}}{r^{4}} + k^{2} \frac{Ae^{ikr}}{r^{2}},\quad\\
	 R^{ij}_{\,, \,ij}+\frac{1}{2}\Delta^{2} g^{T}\lesssim \frac{\mathcal{B}}{r^{6}} + k^{4} \frac{Ae^{ikr}}{r^{2}},\quad\\
	 \Delta R^{ij}_{\,, \,ij}+\frac{1}{2}\Delta^{3} g^{T}\lesssim \frac{\mathcal{B}}{r^{8}} + k^{6} \frac{Ae^{ikr}}{r^{2}}.\quad
	 \end{eqnarray}

	 Then from (\ref{vinculo_HN}) and (\ref{vinculo_HP} )  we obtain the coupled system of two sixth order partial differential equations for  $N$ and $g^{T}$:

	\begin{eqnarray}
	\label{equation_for_gT_and_N_1}
	\left( \beta \Delta  +  \alpha_{1}  \Delta^{2}  +\alpha_{3} \Delta^{3}\right) g^{T} + 2 \left( \alpha \Delta  -  \alpha_{2} \Delta ^{2}+\alpha_{4} \Delta ^{3}\right) N \nonumber \\ 
	\lesssim \frac{\tilde{B}}{r^{2}} + k^{2}P_{2}(k^{2}) \frac{\hat{A}e^{ikr}}{r^{2}}, \qquad \\
	\left[ -\frac{\beta}{2}  \Delta + \frac{1}{2}(3\beta_{1}+8\beta_{2} ) \Delta^{2} - \frac{1}{2}(3\beta_{3}+8\beta_{4}) \Delta^{3} \right] g^{T} \nonumber \\ 
	\label{equation_for_gT_and_N_2}
	-2\left[ \beta \Delta+ \alpha_{1} \Delta^{2}  +\alpha_{3} \Delta^{3}\right] N \nonumber \\
	\lesssim \frac{\hat{B}}{r^{2}} + k^{2}P_{2}(k^{2}) \frac{\tilde{A}e^{ikr}}{r^{2}}.\qquad
	\end{eqnarray}

From (\ref{equation_for_gT_and_N_1}) and (\ref{equation_for_gT_and_N_2}), it follows that the $B$ factor must vanish, and the non-oscillatory contribution is of order $O(1/r^3 )$. It means that the non-oscillatory part of the momentum is of order $O(1/r^2)$. This point is very relevant for the next arguments.

By assumption the fields decay as in (\ref{asymptotic-metric-deviation}) on the wave zone. From the equations  (\ref{equation_for_gT_and_N_1}, \ref{equation_for_gT_and_N_2}) it then follows that the non-oscillatory part of the solution for $g^T$ and $N$ can contribute to this order, since on the right hand side there may be terms of order $1/r^{3}$. On the other side, the oscillatory terms of the form $e^{ikr}/r$ satisfies, on the wave zone
\begin{equation}
(\Delta + k^2) \frac{e^{ikr}}{r}  =0  \,.
\end{equation}
Consequently, for such terms the equations (\ref{equation_for_gT_and_N_1}, \ref{equation_for_gT_and_N_2}), considering terms $g_o^T\frac{e^{ikr}}{r}$ and $N_o\frac{e^{ikr}}{r}$ imply
\begin{eqnarray}
\label{equations_constraints_1}
D_1(k^2) g_o^T + D_2(k^2) N_o=0 \,, \\
\label{equations_constraints_2}
D_2(k^2) g_o^T + D_3(k^2) N_o=0 \,,
\end{eqnarray}
where we define polynomials
\begin{eqnarray}
D_1(k^2)&\equiv& \frac{1}{8}\left[-(3\beta_3+4\beta_4)k^6- (3\beta_1+8\beta_2)k^4-\beta k^2\right]\,, \nonumber\\
\\
D_2(k^2)&\equiv& \frac{1}{2}\left[-\alpha_3 k^6+\alpha_1 k^4-\beta k^2\right]\,, \\
D_3(k^2)&\equiv& -\alpha_4 k^6-\alpha_2 k^4-\alpha k^2\,,
\end{eqnarray}
equations (\ref{equations_constraints_1}, \ref{equations_constraints_2}) yield $g_o^T=N_o=0$  for these zero modes, except when
\begin{equation}
\label{determinante}
D(k^{2})\equiv D_1(k^2)D_3(k^2) - D_2^2(k^2)=0.
\end{equation}
In the case where the zero modes vanish, the behaviour of $g^T$ and $N$ on the wave zone,  taking into account the asymptotic flatness condition $N-1\rightarrow 0$, is
\begin{equation}
	N-1 \sim g^{T} \lesssim\frac{B}{r} +\frac{\hat{A}  e^{ikr}}{r^{2}},
	\end{equation} 
hence,
	\begin{eqnarray}
	a_{i}\lesssim \frac{B_{i}}{r^2}+  k \frac{\hat{A}_{i}  e^{ikr}}{r^{2}}, \\
	\label{a^{1}a_{i}}
	a_{i} a^{i}\lesssim \frac{B}{r^{4}}+ k^{2} \frac{\hat{A}  e^{ikr}}{r^{4}},\\
	\label{nabla^i a_{i}}
	\nabla^{i}a_{i} \lesssim \frac{B}{r^{3}}+k^{2} \frac{\hat{A}  e^{ikr}}{r^{2}}.
	\end{eqnarray} 
On the other side, 
there are a finite number of solutions of (\ref{determinante}) for which there are nontrivial solutions for $g_{0}^T$ and $N_{0}$ and hence contributions of the form $\frac{e^{ikr}}{r}$ for $g^T$ and $N$.  In the low energy case (\ref{determinante}) reduces to a term proportional to $k^2$ with factor $-\frac{\beta^2}{4}(\beta - \frac{\alpha}{2})$ which is different from zero for the values of $\beta$ and $\alpha$ determined from experimental data.  At low energies \cite{mestra2021}, we have always  $g_{0}^T=N_{0}=0$.	
In the 
next section we analyze  the general solution of (\ref{determinante}).
%
\subsection{Constraints resolution}
\label{Constraints resolution}
The polynomial of equation (\ref{determinante}) has $D(k^{2})=k^{4}Q(k^{2})$ where we defined  the quartic polynomial in $k^{2}$ as

\begin{eqnarray}
\label{Q-poli}
Q(k^{2})&\equiv&ak^{8}+bk^{6}+ck^{4}+dk^{2}+e,\\
a&\equiv&    \alpha_{4}\left( 3\beta_{3} + 4  \beta_{4}\right)- 2 \alpha_{3}^{2},\\
b&\equiv&   \alpha_{2}\left( 3 \beta_{3} + 4 \beta_{4}\right) +  \alpha_{4} \left( 3 \beta_{1} + 8  \beta_{2}\right) \nonumber \\&&+ 4 \alpha_{1} \alpha_{3}, \\
c&\equiv&   \alpha \left( 3 \beta_{3} + 4  \beta_{4}\right)  +  \alpha_{2}\left( 3 \beta_{1} + 8  \beta_{2}\right) \nonumber \\ &&+\beta\left(  \alpha_{4}-4 \alpha_{3} \right)- 2 \alpha_{1}^{2},\\
d &\equiv&  \alpha \left( 3 \beta_{1} + 8  \beta_{2}\right) + \beta\left(4 \alpha_{1}  + \alpha_{2} \right),\\
e&\equiv& \beta \left(\alpha  - 2 \beta\right).
\end{eqnarray}

 For the theory to be power counting renormalizable we demand $a\neq 0$ \cite{BellorinRestuccia2016B}.  Experimental tests at  low energies require that $e\leq 0$.  Note that $k^{2}=0$ is a root of  (\ref{determinante}),  this root is not considered in the analysis  in the wave zone due to the  condition $kr\gg1$.
 
 We are interested in the  roots of $Q(k^{2})=0$, we follow the Descartes approach. The change of variable $k^{2}=s-\frac{b}{4a}$  yields the depressed polynomial associate to $Q(k^{2})=0$ 
 
 \begin{eqnarray}
 \label{P(s)-pol}
 P(s)\equiv a\left(s^{4}+ps^{2}+qs
+r\right)=0, \\
p\equiv\frac{c}{a}-\frac{3}{8} \frac{b^{2}}{a^{2}} \,,\\ q\equiv \frac{1}{8} \frac{b^{3}}{c^{3}} +\frac{d}{a}-\frac{bc}{2a^{2}}\,, \\ r\equiv-\frac{3}{256} \frac{b^{4}}{a^{4}}+\frac{1}{16}\frac{cb^{2}}{a^{3}}-\frac{bd}{a^{3}}+\frac{e}{a} \,.
\end{eqnarray}

We  factorized

\begin{eqnarray}
\label{polinomio_P}
 P(s)=a\left(s^{2}+fs+g\right)\left(s^{2}-fs+h\right) \\
 g+h=f^{2}+p,\\
f(h-g)=q,\\
gh=r,
\end{eqnarray}

 where $f^{2}$ satisfies solvent cubic polynomial
 \begin{equation}
 \label{R-poly}
 R(f^{2})\equiv f^{6}+2pf^{4}+(p^{2}-4r)f^{2}-q^{2}=0.
 \end{equation}
There always exist a real positive solution for (\ref{R-poly}). In fact,   if $q^{2} >0$, then (\ref{R-poly}) always has a strictly positive solution. If $q=0$ we can chose $f^{2}=0$ as root and $P(s)$ is a bi-quadratic polynomial its roots can be determined using the general quadratic formula for $s^{2}$. If $f \neq 0$ from (\ref{polinomio_P}) we can calculate $g$ and $h$ via 
 
\begin{eqnarray}
    h&=&\frac{1}{2}\left(f^{2}+p+\frac{q}{f}\right),\\
    g&=&\frac{1}{2}\left(f^{2}+p-\frac{q}{f}\right).
\end{eqnarray}

Finally, we can express (\ref{Q-poli}) in a factorized form 
\begin{equation}
\label{factorizacionk}
    Q(k^2)=a\left(k^{4}+Ak^{2}+B\right)\left(k^{4}+Ck^{2}+D\right).
\end{equation}
where the constants $A, B, C, D$ are related to the coupling parameters by
\begin{eqnarray}
   A=f+\frac{1}{2}\frac{b}{a},\quad B=\frac{1}{4}\frac{bf}{a} +\frac{1}{16}\frac{b^{2}}{a^{2}}+g,\\
   C=-f+\frac{1}{2}\frac{b}{a},\quad  D=-\frac{1}{4}\frac{bf}{a} +\frac{1}{16}\frac{b^{2}}{a^{2}}+h \,.
\end{eqnarray}

If $B$ and $D$ have nonzero imaginary part, then there are not zero modes. If they are real, then there are not zero modes iff the coupling parameters satisfy the following conditions
\begin{equation}
\label{Condicionesacosntantes}
    a<0,\quad B>0, \quad D>0, \quad -2\sqrt{B}<A, \quad-2\sqrt{D}<C.
\end{equation}
The zero modes have, generically, a divergent contribution to the total gravitational  energy. In fact the energy density obtained from the boundary terms in the Hamiltonian, evaluated at the zero modes decay as $1/r$.

We have thus shown that the conditions (\ref{Condicionesacosntantes}) must be satisfied in order to have a consistent formulation of Ho\v{r}ava-Lifshitz gravity.

\subsection{The dynamical equations on the wave zone}
	
	The  T+L components of Equation (\ref{gpunto}) in the gauge $g_{ij,j}=0$, becomes
		\begin{eqnarray}
		\label{gpinto-TT}
		\dot{g}_{ij}^{TT}-2 \pi_{ij}^{TT}+ \left(\mu g_{ij}\right)^{TT} &\lesssim&\frac{B}{r^{2}}+k^{2}\frac{\hat{A}  e^{ikr}}{r^{2}},\\
		\label{gpunto-L}
		-2 N_{(i,j)} + \left(\mu g_{ij}\right)^{L} &\lesssim&\frac{B}{r^{2}}+k^{2}\frac{\hat{A} e^{ikr}}{r^{2}},\\
		\label{gpunto-T}
		\dot{g}_{ij}^{T}+\left(\mu g_{ij}\right)^{T}&\lesssim&\frac{B}{r^{2}}+k^{2}\frac{\hat{A}  e^{ikr}}{r^{2}},
	\end{eqnarray}
	where $\mu g_{ij}=  \left(\mu g_{ij}\right)^{TT}+\left(\mu g_{ij}\right)^{T}+\left(\mu g_{ij}\right)^{L}$ and $\left(\mu g_{ij}\right)^{T}=\mu \delta_{ij}-\frac{1}{\Delta}\mu_{,ij}+\mathcal{O}\left(1/r^{2}\right)$, $\left(\mu g_{ij}\right)^{L}=\frac{1}{\Delta}\mu_{,ij}+\mathcal{O}\left(1/r^{2}\right)$ and 
	$\left(\mu g_{ij}\right)^{TT}=\mathcal{O}\left(1/r^{2}\right)$. Then from (\ref{gpunto-T}) we get $\mu\lesssim\frac{B}{r^{2}}+k^{2}\frac{\hat{A}  e^{ikr}}{r^{2}}$   and consequently
	 \begin{equation}
	\label{canonica-gpinto-TT}
	\dot{g}_{ij}^{TT}-2 \pi_{ij}^{TT} \lesssim\frac{B}{r^{2}}+k^{2}\frac{\hat{A} e^{ikr}}{r^{2}} \,,
	\end{equation}
	\begin{equation}
	\label{gpunto_(i,j)}
	N_{(i,j)}\lesssim\frac{B}{r^{2}}+k^{2}\frac{\hat{A}  e^{ikr}}{r^{2}} \,. 
	\end{equation}
	We then have $N_{i}\lesssim \frac{B}{r}+k \frac{\hat{A}  e^{ikr}}{r^{2}}$.

	Besides we get the following   contributions  in  (\ref{pipunto}) 
	\begin{eqnarray}
	\beta N\frac{\delta R}{\delta g_{ij}}	
	&=&\frac{\beta}{2}\Delta g^{T}+\mathcal{O}(1/r^ {2}) \,,\\
	\beta_{1} N\frac{\delta}{\delta g_{ij}}(R_{ij} R^{ij})	
	&=& \frac{\beta_{1}}{2}\Delta^{2}g_{ij}^{TT}+\mathcal{O}(1/r^ {2}) \,,\\
	\beta_{3} N\frac{\delta}{\delta g_{ij}}(\nabla_{i}R_{jk} \nabla^{i}R^{jk})
	&=& -\frac{\beta_{3}}{2}\Delta^{3}g_{ij}^{TT}+\mathcal{O}(1/r^ {2}) \,.\qquad
	\end{eqnarray}
	Then  we obtain,
	
	\begin{equation}
	\label{canonical-pi-punto-TT}
	\dot{\pi}_{ij}^{TT}=\frac{1}{2} \left(\beta \Delta+\beta_{1}\Delta^{2}-\beta_{3}\Delta^{3}\right)g_{ij}^{TT}+\mathcal{O}(1/r^{2}) \,.
	\end{equation}
	
	Equations (\ref{canonica-gpinto-TT}) and  (\ref{canonical-pi-punto-TT}) show that in the wave zone the TT-part of the fields up to order $\mathcal{O}(1/r)$ satisfies the sixth order  partial differential equation
	\begin{eqnarray}
	\label{wave-six-gpuntoTT}
	\ddot{g}_{ij}^{TT}-\left(\beta \Delta+\beta_{1}\Delta^{2}-\beta_{3}\Delta^{3}\right)g_{ij}^{TT}&=&0,\\
	\label{wave-six-ppuntoTT}
	\ddot{\pi}_{ij}^{TT}-\left(\beta \Delta+\beta_{1}\Delta^{2}-\beta_{3}\Delta^{3}\right)\pi_{ij}^{TT}&=&0.
	\end{eqnarray}
	The terms with spatial derivative greater than 2, are due to the contributions of the potential terms $\mathcal{V}^{(2)}$ and $\mathcal{V}^{(3)}$.  
	
	We can obtain the low energy limit of Equations (\ref{wave-six-gpuntoTT}) and (\ref{wave-six-ppuntoTT}) if we assume that $k_{max}\ll K_{UV}$, where $K_{UV}$ is the ultra-violet cut-off. Then the Laplacian acting on the TT modes is the leading term among the spatial derivatives ones. In this case (\ref{wave-six-gpuntoTT}) and (\ref{wave-six-ppuntoTT}) reduce to the wave equation for the TT modes 
	\begin{eqnarray}
	\ddot{g}^{ijTT}-\beta \Delta g^{ijTT}&=&0 \,,\\
	\ddot{\pi}^{ijTT}-\beta \Delta\pi^{ijTT}&=&0 \,.
	\end{eqnarray}
	
	The canonical modes travel with a speed $\sqrt{\beta}$,  and it only differs from the wave zone of General Relativity by the value of the velocity of propagation. Gravitational Wave experiments restrict this value to be between $-3\times {10}^{-15}$ and $+7\times {10}^{-16}$ times the speed of light \cite{Abbott_2017,Abbott2019b}.

\section{Solution of the Ho\v{r}ava-Lifshitz wave equation at the kinetic conformal point}
The solution to  (\ref{wave-six-gpuntoTT}) can be written in terms of the spherical Hankel functions $h_{l}^{(1)}(kr)$ and $h_{l}^{(2)}(kr)$. In fact,
\begin{equation}
\label{solution_u}
u(r,k) \equiv  \sum_{l,m} \left[A_{lm}^{(1)} h_{l}^{(1)}(kr) + A_{lm}^{(2)} h_{l}^{(2)}(kr) \right] Y_{lm}(\theta, \phi) \,,
\end{equation}
satisfies
\begin{equation}
\label{equation_for_solution_u}
-\Delta u(r,k) = k^2 u(r,k) \,,
\end{equation}
hence

\begin{eqnarray}
\label{solution_u_2}
\left[\beta (-\Delta) - \beta_1 (-\Delta)^2 - \beta_3 (-\Delta)^3 \right] u(r,k)\nonumber \\ = \left(\beta k^2 - \beta_1 k^4 - \beta_3 k^6 \right) u(r,k) \,.
\end{eqnarray}

Finally, $\Psi (r,t) = e^{i\omega t }u(r,k)$ satisfies eq. (\ref{wave-six-gpuntoTT}) provided
\begin{equation}
\label{solution_u_3}
\beta k^{2} - \beta_{1} k^{4} - \beta_{3} k^{6}  = \omega^{2} \,.
\end{equation}
For a given $l$, 
\begin{equation}
\label{solution_u_4}
h_{l}^{(1)}(kr) = \left[(-i)^{l+1} \sum_{s=0}^l \frac{i^s}{s! (2kr)^s} \frac{(l+s)!}{(l-s)!}\right] \frac{e^{ikr}}{r}
 \,,
\end{equation}
and $h_{l}^{(2)}(kr)=\overline{h_{l}^{(1)}}(kr)$ where $\overline{h}$ denotes complex conjugation.

In the wave zone 
\begin{equation}
\label{solution_u_5}
h_{l}^{(1)}(kr) \approx (-i)^{l+1}  \frac{e^{ikr}}{r} \,, \quad h_{l}^{(2)}(kr) \approx (i)^{l+1}  \frac{e^{-ikr}}{r} \,,
\end{equation}
hence

\begin{eqnarray}
\label{solution_u_6}
u(r,k)= \sum_{l,m} \left[A_{lm}^{(1)} (-i)^{l+1}   Y_{lm}(\theta, \phi)\right]\frac{e^{ikr}}{r} \nonumber \\+ \sum_{l,m}\left[A_{lm}^{(2)}(i)^{l+1} Y_{lm}(\theta, \phi)  \right] \frac{e^{-ikr}}{r}\,.
\end{eqnarray}

Finally, the outgoing solution to (\ref{wave-six-gpuntoTT}) in the wave zone is
\begin{equation}
\label{solution_u_8}
\Psi = \frac{e^{-i \omega t + ikr}}{r} \sum_{l,m} A_{lm}^{(1)} (-i)^{l+1}   Y_{lm}(\theta, \phi)\,.
\end{equation}
where $k$ is the solution of the dispersion relation (\ref{solution_u_3}).
In order to have a positive left hand member of (\ref{solution_u_3}) the couplings must satisfy  \cite{BellorinRestuccia2016B}
\begin{equation}
\label{XXX}
\alpha \ne 2 \beta \,, \quad \beta > 0 \, , \quad \beta_3 < 0 \,, \quad \beta_1 < 2 \sqrt{\beta |\beta_3|}.   
\end{equation}
%
%
%
In this case given $\omega^2$ there always exists a unique $k^2$ satisfying the dispersion relation. Consequently, given $\omega$ there exists an outgoing solution with dependence on the radial coordinate $  {e^{-i \omega t + ikr}}/{r}$.
The relation between $\omega$ and $k$ can be rewritten as
\begin{equation}
\label{solution_u_10}
\frac{\omega^2}{k^2} = \beta - \beta_1 k^2 - \beta_3 k^4 \,, 
\end{equation}
it tells us that at low energies the speed of propagation of the gravitational wave is $\sqrt{\beta}$ while at high energies the speed depends on the energy of the wave. Under the above assumptions on the coupling constants, the right hand member of (\ref{solution_u_10}) is always positive.

\section{Discussion}
We showed that in the Ho\v{r}ava-Lifshitz theory at the kinetic conformal there exists a wave zone on which the physical degrees of freedom, the transverse-traceless (TT) tensorial modes propagate as free waves. On that region of space, provided the set of restrictions on the coupling parameters (\ref{Condicionesacosntantes}) is assumed, the other components of the metric on the spacelike leaves of the foliation as well as the lapse function, are of the same order in powers of $1/r$, $r$ being the distance to a bounded source. They constitute a static Newtonian background which does not interact with the propagation of the wave modes.
At larger distance from the source than the wave zone the TT components of the metric decay with higher powers of $1/r$, while the T modes vary in a way that the energy of the system is finite.
The set of restrictions, equation (\ref{Condicionesacosntantes}), is a necessary and sufficient condition for the non-existence of zero modes to the solutions of the constraints. 
The zero modes would have a divergent contribution to the gravitational energy and must be avoided. The wave solution describing the propagation of the physical degrees of freedom satisfies a linear partial differential equation quadratic on time derivatives but cubic on the Laplace operator. At low energies, the infrared limit of the renormalization flow of the Ho\v{r}ava-Lifshitz theory, the equation reduces to the wave equation, with a speed of propagation characterized by the square root of a coupling constant, which should take at this limit a value very near to the speed of light $c$. At higher energy level the high spacelike derivative terms in the potential of the Hamiltonian become relevant and the wave solutions are characterized by a dispersion relation which means that the speed of propagation of the physical modes becomes energy dependent. The solution of the dispersion relation requires additional restrictions on the coupling parameters, they must satisfy (\ref{XXX}).

The Ho\v{r}ava-Lifshitz theory at the kinetic conformal point in the low energy limit depends on two coupling constants $\beta$ and $\alpha$, which may have a bounded range of values in order that the predictions of the theory become consistent with the known experimental data, for $\beta =1$ and $\alpha=0$ at low energies the theory is the same as GR in a particular gauge. For $\alpha \neq0$ 
the static solution with spherical symmetry does not describe a black hole but a throat connecting an asymptotic flat space and a space with an essential singularity  \cite{BellorinRestuccia2016B}. 
That is, the physics of the theory is different from GR, as predicted in \cite{Horava2009}, but in the limit $\beta \rightarrow 1$  and $\alpha \rightarrow 0$  the geometry external to the throat is very similar to the geometry external to the black hole of the Schwarzschild solution in GR, however only in the limit $\beta =1$ and $\alpha =0$ there is a black hole. An analogous relation to GR occurs with the gravitation radiation. We have shown that the gravitation radiation predicted by the theory at low energies has similar properties as the radiation in GR, but at higher energies the wave solutions satisfy a dispersion relation characteristic of the Ho\v{r}ava-Lifshitz theory. 
The higher order  spacelike derivatives terms in the evolution equation are related to the higher order derivative terms in the potential that determine, at the UV regime, the power counting renormalizability of the theory, the main virtue of the Ho\v{r}ava proposal.

An interesting study would be to analyze the radiative properties of the gravitation-electromagnetic interaction on the wave zone. It is known that the gravitation and electromagnetic degrees of freedom propagate, in a perturbative approach, with the same speed but their interaction behavior in the wave zone is unknown \cite{BellorinRestucciaTello2018b, RestucciaTello2020}.
	
\begin{acknowledgements}
J. Mestra-Páez acknowledge financial support from  Beca Doctorado Nacional 2019 CONICYT, Chile.   N° BECA: 21191442. J. M. P is supported by the projects ANT1956 and SEM18-02 of the University of Antofagasta, Chile. 
\end{acknowledgements}
\bibliographystyle{elsarticle-num}
\bibliography{REF}

\end{document}